\documentclass[12pt,a4paper,oneside]{article}
\pdfoutput=1
\usepackage[a4paper]{geometry}
\usepackage{amsmath,amssymb,amsfonts,amsxtra, mathrsfs, makeidx,graphics,graphicx,amsthm,epsfig, bm,longtable,float, color,tikz,mathtools,xfrac,footnote,rotating, lscape, makecell, environ,mathtools, empheq}

\usepackage{jheppub}

\usepackage{subfig}
\usepackage{multirow}
\usepackage{adjustbox}

\graphicspath{{fig/}}

\usepackage{amsthm}
\usepackage{hyperref}

\usepackage{commath}
\usepackage[english]{babel}
\usepackage{setspace}

\usepackage{verbatim,booktabs}

\usepackage{latexsym}

\DeclareMathOperator{\Tr}{Tr}

\newcommand{\pt}{\widetilde{\phi}}
\renewcommand{\Re}{\text{Re}}
\renewcommand{\Im}{\text{Im}}

\title{Kerr-Newman black holes from $\mathcal{N}=1^*$}

\author[a]{Antonio Amariti}
\author[a,b]{and Alessia Segati}

\affiliation[a]{INFN, Sezione di Milano, Via Celoria 16, I-20133 Milano, Italy}
\affiliation[b]{Dipartimento di Fisica, Università degli studi di Milano, Via Celoria 16, I-20133, Milano, Italy}

\emailAdd{antonio.amariti@mi.infn.it}
\emailAdd{alessia.segati@mi.infn.it}

\abstract{The microstate counting of charged rotating AdS$_5$ supersymmetric black holes has been reformulated in terms of an extremization problem, obtained from the superconformal index of the 4d dual SCFT.
On the gravitational side this  problem corresponds to the attractor mechanism of the theory KK reduced on AdS$_4$.
 Such procedure has indeed been successfully applied to some consistent truncations with a known field theory dual description.
 In this paper we study the case of the Leigh-Strassler fixed point along these lines, finding an agreement between the field theory and the gravitational results.
 }

\begin{document}

\maketitle

\section{Introduction}

Attractor mechanisms in supergravity are related through the AdS/CFT 
correspondence to  extremization problems
 in the dual superconformal field theories.
Many explicit cases have been worked out in the recent past.
An early attempt was provided in \cite{Tachikawa:2005tq} where it was shown that
the maximization of the conformal anomaly $a_{4d}$ \cite{Intriligator:2003jj} corresponded to the minimization
of the scalar potential in the AdS$_5$ supergravity dual.
In this case the R-charges were related to the scalar fields in the vector multiplets,
and the role of the hypermultiplets was discussed as well (see \cite{Szepietowski:2012tb} for an explicit example).
In a related paper \cite{Barnes:2005bw} it was shown that a similar relation  holds between the 
attractor mechanism and the coefficient of the two-point function for the R-current
$\tau_{RR}$ (see \cite{Barnes:2005bm} for an explicit derivation of $\tau_{RR}$-minimization). The latter result is interesting because it can be extended
to other dimensions.
For example the extremization of $\tau_{RR}$ in $2d$, corresponding to the central charge,
led to the principle of $c$-extremization \cite{Benini:2012cz}. 
The supergravity dual mechanism was 
 discussed in  \cite{Benini:2013cda,Karndumri:2013dca,Karndumri:2013iqa,Benini:2015bwz,Amariti:2016mnz}.
 It has been shown that $c$-extremization  can be reformulated in terms of  
an attractor mechanism in AdS$_3$. 

Furthermore in three dimensions $\tau_{RR}$ coincides with the free energy  
on $S^3$ for holographic theories \cite{Closset:2012vg}, 
and this allows to relate the attractor mechanism 
to localization \cite{Freedman:2013oja,Bobev:2018wbt} (see also \cite{Amariti:2015ybz} for a discussion on $\tau_{RR}$ minimization in  AdS$_4$ gauged supergravity).

By generalizing this idea in \cite{Benini:2015eyy,Hosseini:2016tor,Hosseini:2016ume}  it was shown that the extremization of another 
quantity obtained from localization, the topologically twisted index \cite{Benini:2015noa}, can be associated 
at large $N$ to the entropy of AdS$_4$ BPS black holes.
This mechanism has been then generalized to other dimensions and associated to the attractor mechanism
\cite{Benini:2016rke,Hosseini:2016cyf,Bobev:2017uzs,Benini:2017oxt,Hosseini:2017fjo,Cabo-Bizet:2017xdr,Bobev:2018uxk,Hosseini:2018uzp,Hosseini:2018dob,Hosseini:2018usu}.

A more recent extremization problem on the field theory side regards the extremization of the
entropy function and its relation with the entropy of 5d rotating black holes.
It has been shown in \cite{Hosseini:2017mds} that the Legendre transform of such function, for the case of
$\mathcal{N}=4$ SYM, gives rise to the entropy of the 
electrically charged rotating BPS black holes in AdS$_5 \times S^5$
\cite
{Gutowski:2004yv,Gutowski:2004ez,Chong:2005da,Chong:2005hr,Kunduri:2006ek}.
On the gravitational side this quantity has been shown to originate from the on--shell action of the 
Euclidean black hole \cite{Cabo-Bizet:2018ehj}, while on the field theory side 
 the  entropy function has then been shown to originate from 
the superconformal index \cite{Choi:2018hmj,Benini:2018ywd}.

However the  dual attractor mechanism for rotating five-dimensional black holes in gauged supergravity 
 is unknown and for obtaining the gravitational  function a more sophisticated construction has been necessary \cite{Hosseini:2017mds},
based on the  general reduction of BPS attractors in gauged supergravity performed in \cite{Hristov:2014eza}.
In this case, after  fixing the two angular momenta to be equal, it is possible to dimensionally reduce the five-dimensional solution down to four dimensions along the $U(1)$  Hopf fiber of the enhanced $SU(2)\times U(1)$ isometry of the black hole metric on the squashed sphere. In this framework 
the dual extremization problem was
formulated in terms of the four-dimensional BPS black hole attractor mechanism \cite{DallAgata:2010ejj}. 
More recently this result has been extended to 
the case of truncations with hypermultiplets, and checked for the case of $T^{1,1}$ studied in \cite{Benini:2020gjh}
and of M5 branes in \cite{Hosseini:2020mut}.
Observe that in these last cases the results are obtained by conjecturing the existence of a BPS Kerr-Newman black hole,
that has nevertheless never been directly studied from a 5d analysis, differently from the case of $S^5$.
By using this assumption on the existence of the 5d black hole the final results on its entropy, obtained from the AdS$_2 \times S^2$ attractor, have been matched with the expectations form the superconformal index.

Motivated by that construction,
in this note we discuss the relation between the  superconformal index of the 
$\mathcal{N}=4$ theory with superpotential $W=\epsilon_{ijk} \Phi_i \Phi_j \Phi_k$ perturbed by the $\mathcal{N}=1$ Leigh--Strassler (LS) deformation, $\Delta W = m \Phi_3^2$, and the entropy of the holographic dual BPS Kerr--Newman rotating black hole.

We consider the truncation of \cite{Ceresole:2001wi} corresponding to  a model with  one vector multiplet  and one hypermultiplet, and a $U(1) \times U(1)$ gauging of the isometries of the scalar manifold.
This truncation is less rich that the one studied in \cite{Benini:2020gjh}, because only the R-symmetry current is captured, but it has other interesting properties that require a detailed analysis.
The first non trivial aspect is that we have to apply a local rotation that aligns the Killing prepotentials. This allows 
us to use the approach of  \cite{Hosseini:2017mds,Benini:2020gjh} in the framework of general matter coupled to 
$\mathcal{N}=2$ gauged supergravity.
A second aspect is related to the different conventions in the original truncations of \cite{Ceresole:2001wi} with respect to the ones used in \cite{Hosseini:2017mds,Benini:2020gjh}. This amounts to a slightly different choice of the Kaluza-Klein (KK) ansatz in our case. Motivated by this difference here we perform a general analysis of the KK ansatz.

In this setup we then reduce the 5d gauged supergravity to 4d and study the AdS$_2$ attractor of the conjectured black hole solution. This gives rise to an extremization problem that corresponds to the one found in \cite{Amariti:2020jyx} from the saddle point analysis of the superconformal index, once restricted to the fugacities visible in the truncation of \cite{Ceresole:2001wi}. Here we further corroborate the result by following the approach of \cite{Benini:2018ywd}, using the solutions of the Bethe Ansatz Equations (BAEs).

The paper is organized as follows. In section \ref{LSrev} we review the basic aspects of the 5d truncation dual to the LS fixed point discussed in \cite{Ceresole:2001wi}.
In section \ref{genattr} we review the 5d/4d reduction along the lines of \cite{Hosseini:2017mds,Benini:2020gjh}
and we apply the construction to the truncation of the LS fixed point. We further find the extremization problem 
that originates from the AdS$_2$ attractor.
This problem is then matched with the field theory expectations in section \ref{FTHOLM}, where we also derive the 
entropy function of the conjectured black hole through the analysis of the BAE. Then in section \ref{concl}
we give some conclusive remarks.
We also added some appendices in order to fix notations on 5d \ref{5dt} and 4d \ref{4dt} $\mathcal{N}=2$ gauged supergravity. As anticipated we also give a detailed analysis of the KK ansatz used in the bulk of the paper 
in appendix  \ref{KK}.

\section{A consistent truncation dual to the LS fixed point}
\label{LSrev}
In this section we review the relevant aspects of the 
AdS$_5$ gauged supergravity dual of the 4d SCFT LS fixed point.
The flow to such an IR fixed point was first reproduced from the holographic perspective in \cite{Freedman:1999gp}
within $\mathcal{N} = 8$ supergravity.
It was then shown that it could be obtained in  $\mathcal{N} = 2$ gauged supergravity in \cite{Ceresole:2001wi}.
The starting point in the analysis of \cite{Ceresole:2001wi} is a model with  one vector multiplet  and one hypermultiplet, and a $U(1) \times U(1)$ gauging of the isometries of the scalar manifold $ \mathcal{M} =  \mathcal{SM}  \times \mathcal{QM} $
(see Appendix \ref{5dt} for a review of 5d $\mathcal{N}=2$ gauged supergravity). 
The UV and the IR fixed points are connected by an R-symmetric flow along 
the quaternionic-K\"ahler manifold $\mathcal{QM}$.
The IR fixed point corresponds to the LS fixed point and it is the starting point of our analysis.
In the following we briefly review the relevant aspects of such IR fixed point, 
referring the reader to \cite{Ceresole:2000jd} for conventions on the formalism of 
general matter coupled $\mathcal{N}=2$ gauged supergravity
in five dimensions.
We also  discuss an useful
manipulation of the results of \cite{Ceresole:2001wi}, corresponding to an $Sp(1)$  rotation
\footnote{This $Sp(1)$ is the one appearing in the $Sp(n_H) \times Sp(1)$ holonomy group of $\mathcal{QM}$. It is ofter referred to as an $SU(2)_R$ symmetry.} of the Killing prepotentials, that becomes relevant in the study of the 5d/4d reduction performed below.

\subsubsection*{The model}

The model considered in \cite{Ceresole:2001wi} to reproduce the FGPW flow consists of
5d $\mathcal{N}=2$ gauged supergravity with one vector ($n_V=1$) and one hypermultiplet ($n_H=1$), with scalar manifold
\begin{equation}
  \label{scalarmanifold:hyper}
  \mathcal{M} = O(1,1) \times \frac{SU(2,1)}{SU(2) \times U(1)}.
\end{equation}
The scalar manifold of the very special geometry is specified by the totally symmetric tensor $C_{IJK}$ as
\begin{equation}
  \label{scalarmanifold:special}
  \mathcal{SM} = \bigl\{ \mathcal{V}(h) \equiv  C_{IJK} h^I h^J h^K =1 \bigr\},
\end{equation}
where $h^I \ (I = 1,2)$\footnote{
For reasons related to the 5d/4d analysis  here we shift the index $I=0,\dots,1$ by one unit,
i.e. $I=1,\dots,n_V+1.$} are coordinates on $\mathbb{R}^2$ and $h^I(\phi)$ represent the sections of the special geometry. In this truncation we have
\begin{equation}
  \label{C122}
  C_{122}=3\sqrt{3},
\end{equation}
and thus
\begin{equation}
  \label{SM:our}
  \mathcal{V}(h) = 9\sqrt{3} h^1 (h^2)^2.
\end{equation}
The quaternionic-K\"ahler manifold $\mathcal{QM}$ is parametrized by four real scalars $q^u = \{ V, \sigma, \theta, \tau \}$, with $V >0$. The metric is given by
\begin{equation}
  \label{MQ:metric}
  ds^2 = \frac{dV^2}{2V^2} + \frac{1}{2V^2}\bigl(d\sigma - 2\tau \, d\theta + 2\theta \, d\tau \bigr)^2 + \frac{2}{V} (d\theta^2 + d\tau^2 ).
\end{equation}

\subsubsection*{The gauging}
The Killing vectors are given by a linear combination of the generators of the $SU(2)$ and $U(1)$ subgroups of the $SU(2,1)$ isometry group of the metric \eqref{MQ:metric},  denoted as $T_3$ and $T_8$ in \cite{Ceresole:2001wi}. 
The $U(1)\times U(1)$ gauging corresponds to consider the two Killing vectors
\begin{equation}
  \label{killing:trunc}
    K_1 = \frac{3}{\sqrt{2}} \Bigl(T_3 + \sqrt{3} \, T_8 \Bigr), \qquad K_2 = \sqrt{3} \Bigl( \sqrt{3} \, T_3 - T_8 \Bigr),
  \end{equation}
  where
  \begin{equation}
    \label{killing:T}
    T_3 = \frac{1}{4} (k_1 + k_6 - 3k_4), \qquad T_8 = \frac{\sqrt{3}}{4} (k_4 + k_1 + k_6)
  \end{equation}
  are the Abelian generators of $SU(2) \times U(1)$, constructed from the eight Killing vectors of the $SU(2,1)$ isometry group of the metric (\ref{MQ:metric}). Explicitly, the relevant generators in (\ref{killing:T}) are
  \begin{equation}
    \label{killing:k}
    \vec{k}_1 =
    \begin{pmatrix}
      0 \\
      1 \\
      0 \\
      0
    \end{pmatrix},
    \quad \vec{k}_4 =
    \begin{pmatrix}
      0 \\
      0 \\
      - \tau \\
      0
    \end{pmatrix},
    \quad \vec{k}_6 =
    \begin{pmatrix}
      2V \sigma \\
      \sigma^2 - (V + \theta^2 + \tau^2 )^2 \\
      \sigma \theta - \tau (V + \theta^2 + \tau^2 )\\
      \sigma \tau + \theta (V + \theta^2 + \tau^2 )
    \end{pmatrix}.
  \end{equation}
  The Killing prepotentials associated to \eqref{killing:trunc} are given by
  \begin{equation}
    \label{prepotentials:trunc}
    \vec{P}_1 =
    \begin{pmatrix}
      -\frac{3(-V \theta + \theta^3 + \sigma \tau + \theta \tau^2)}{\sqrt{2V}} \\
      \frac{3 (V \tau - \theta^2 \tau + \theta \sigma - \tau^3)}{\sqrt{2V}} \\
      - \frac{3(1 + V^2 + \sigma^2 - 6V (\theta^2 + \tau^2) + (\theta^2 + \tau^2)^2)}{4 \sqrt{2} V}
    \end{pmatrix},
    \quad \vec{P}_2 =
    \begin{pmatrix}
      \frac{3 \theta}{\sqrt{V}}\\
      \frac{3 \tau}{ \sqrt{V}} \\
      \frac{3(-V + \theta^2 + \tau^2)}{2V}
    \end{pmatrix}.
  \end{equation}

  \subsubsection*{The FGPW flow}

By choosing the parametrization
\begin{equation}
\label{paramflow}
\sigma=0, \quad V = 1- \theta^2-\tau^2 \equiv 1- \zeta^2  \equiv 1- \tanh^2 \chi  
\end{equation}
with $0<V \leq 1$, $-1<\zeta<1$ and $-\infty < \chi<\infty$
one can reproduce the FGPW flow from the truncation discussed so far.
The fields $\theta$ and $\tau$ can be parameterized as $\theta = \zeta \cos \phi$ and 
$\tau = \zeta \sin \phi$.\\
The UV vacuum corresponds, without loss of generality, to 
\begin{equation}
\sigma=\zeta^2=0,\quad V=1,\quad \rho=1,
\end{equation}
where $\rho$ is the vector modulus. This is the starting point of the flow, holographically dual to $\mathcal{N}=4$ SYM.
In this UV fixed point the $U(1)$ symmetry gauged by the graviphoton is generated by the Killing vector 
$\frac{1}{\sqrt{3}}(K_{1}+K_{2})$. The other massless vector gauges the $U(1)$ isometry associated to $\frac{1}{\sqrt{3}}(K_{1}-K_{2})$. The first $U(1)$ is associated to the $U(1)_R \subset SU(4)_R$ R-symmetry group of $\mathcal{N}=4$ assigning charges $R=\frac{2}{3}$ to the three chiral adjoints in the field theory dual. The other $U(1)$ corresponds to  one of the two Abelian flavor symmetries in the Cartan of $SU(4)_R$.\\
The IR vacuum corresponds, without loss of generality, to 
\begin{equation}
\sigma=0,\quad \zeta^2=\frac{1}{4}, \quad V=\frac{3}{4},\quad \rho=2^{1/6}.
\end{equation}
The flat direction associated to $\phi$ is a marginal deformation from the SCFT dual.
In this IR fixed point the $U(1)$ symmetry gauged by the graviphoton is generated by the Killing vector 
$\frac{2^{2/3}}{\sqrt{6}}(\sqrt 2 K_{1}+K_{2})$. The other  vector in this case is massive, and the broken
isometry associated is generated by $\frac{2^{2/3}}{\sqrt{6}}(\frac{1}{\sqrt{2}} K_2-2 K_1)$.

Summarizing, the spectrum of vector fields and Killing vectors they couple to, around IR fixed point, is given by
  \begin{equation}
    \label{spectrum:IR}
    \begin{aligned}
    A^R_\mu & \equiv \frac{2^{1/3}}{\sqrt{6}} \Biggl( \frac{A_\mu^1}{\sqrt{2}} + 2 \, A_\mu^2 \Biggr) \ : \ m^2 = 0, &  K_R & = \frac{2^{2/3}}{\sqrt{6}} \bigl( \sqrt{2} K_1 + K_2 \bigr), \\
    A^W_\mu & \equiv \frac{2^{1/3}}{\sqrt{6}} \biggl( \sqrt{2} A_\mu^2 - A_\mu^1 \biggr)  \ : \ m^2 = 6 \cdot 2^{4/3} g^2,\quad  &  K_W & = \frac{2^{2/3}}{\sqrt{6}} \Biggl( \frac{K_2}{\sqrt{2}} - 2K_1 \Biggr).
    \end{aligned}
  \end{equation}
  The vector $A^W_\mu$ acquires a mass eating the St\"uckelberg scalar $\chi$. The mass eigenstates are
  \begin{equation}
    \label{mass:eigen}
    \mathbb{B}^I_{\ J} A^J_\mu , \qquad \text{where} \qquad \mathbb{B}= \frac{2^{1/3}}{\sqrt{6}}
    \begin{pmatrix}
      \frac{1}{\sqrt{2}} & 2 \\
      -1 & \sqrt{2}
      \end{pmatrix}
    \end{equation}
    is the matrix that diagonalizes them.
  
  \subsubsection*{Rotating the Killing prepotentials}

The Killing prepotentials $P_I^r$ in (\ref{prepotentials:trunc})  are aligned along $r=3$ at the UV vacuum, while they are misaligned in the IR.
In order to simplify the analysis below here we consider an $Sp(1)$ rotation of these Killing prepotentials such that 
they are aligned around the IR fixed point.
  Following \cite{Amariti:2021cpk}, the $SU(2)$ matrix
  \begin{equation}
    \label{SU2}
    \begin{split}
      U(\zeta,\phi)  & = \exp \biggl( -\frac{i}{2} \sigma^1 f(\zeta) \biggr) \cdot \exp \biggl( - \frac{i}{2} \sigma^3 g(\phi) \biggr),\\
       \text{with } f(\zeta) & = \arctan \biggl( \frac{2 \zeta \sqrt{1-\zeta^2}}{2\zeta^2-1}\biggr), \quad g(\phi) = -\phi + \frac{\pi}{2},
    \end{split}
  \end{equation}
  rotates the moment maps \eqref{prepotentials:trunc} into a form such that the prepotentials become
    \begin{equation}
  \label{Prot}
 \vec P_1=
\left(
\begin{array}{c}
 -\frac{3 }{\sqrt{2}} \sigma  \sinh \chi  \\ 
 \frac{3}{2 \sqrt{2}}  \sigma ^2 \sinh \chi  \\
  \frac{3 }{4 \sqrt{2}} \left(\sigma ^2+2\right) \left(2 - \cosh ^2\chi \right)
 \end{array}
\right),
\quad
 \vec P_2=
 \left(
\begin{array}{c}
 0 \\ 
 0 \\ 
 \frac{3}{2}  \cosh ^2\chi 
\end{array}
\right)
\end{equation}
  where we used (\ref{paramflow}) and parameterized $\theta=\tanh\chi \cos \phi$ and  $\tau=\tanh\chi \sin \phi$. 

The absence of a $\phi$ dependence in the $\vec{P}_I$ signals the fact that there this field is related to a  marginal direction, both at the UV fixed point and at the IR one. 
Furthermore observe that around the IR vacuum, where $\sigma=0$ both $\vec P_1$ and $\vec P_2$ are 
in the form $ P_I^r = \delta_{3r} P_I^3$.  This last observation is valid also if we keep the  $\chi$ dependence explicit.
As observed in similar AdS$_5$ and AdS$_4$ context indeed  the field $\chi$ can be  interpreted as a Lagrange multiplier in the superpotential, enforcing the constraints from the massive vectors on the special geometry. We will see its explicit role
also in our case studied in detail below.

\section{5d/4d reduction and attractor mechanism}
\label{genattr}

Here we briefly illustrate the procedure introduced in \cite{Hosseini:2017mds}.
This amounts of studying a putative 5d black hole by 
relating the near horizon region through a KK dimensional reduction along the
Hopf fiber of the $S^3$ horizon. By considering the limit with 
equal angular momenta the problem can be  
reformulated as the study of a 4d static black hole, and the entropy is obtained through the attractor mechanism. 
In the original construction of \cite{Hosseini:2017mds} only vector multiplets where considered. 
Afterward hypermultiplets have been added to the analysis in \cite{Benini:2020gjh}.\\

The first step consists in reducing the 5d theory on a circle corresponding to the Hopf fiber of $S^3$. This can be achieved by employing the KK reduction ansatz \cite{Benini:2020gjh, Hosseini:2017mds, Klemm:2016kxw, Duff:1986hr}
  \begin{equation}
    \label{ansatz:reduction}
    \begin{split}
      ds^2_{(5)} & = e^{2\pt} ds^2_{(4)} + e^{-4\pt}(dy-A^0_{(4)})^2,\\
      h^I & = - \sqrt{\frac{2}{3}} e^{2\pt} \, \Im \, z^I \qquad (I=1, \dots, n_V+1),
    \end{split} 
  \end{equation}
  where $y$ is the direction of the circular fiber and $A^0_{(4)}$ is the KK vector. Using the constraint on $\mathcal{SM}$ in \eqref{scalarmanifold:special}, the field $\pt$ can be eliminated with $e^{-6\pt}=- \frac{4}{3\sqrt{6}}\mathcal{V}(\Im \, z^I)$.\\

   Notice that our ansatz differs from the one in \cite{Benini:2020gjh} by a factor $\sqrt{\frac{2}{3}}$ in the reduction of the coordinates $h^I$. This is because the 5d Lagrangian of \cite{Ceresole:2001wi} we start with (see \eqref{L:5d}) is written with different conventions with respect to the ones used in \cite{Hosseini:2017mds, Benini:2020gjh}. Such difference requires a modification of the KK ansatz in order to produce a 4d Lagrangian suitable for the analysis of the AdS$_2$ attractor of \cite{DallAgata:2010ejj}. The detailed computation is quite long and is very similar to the one performed in \cite{Benini:2020gjh}. For this reason we report it in Appendix \ref{KK}.\\

The coordinate $y$ is compactified on a circle of length $4\pi/g$ where $g$ is the gauge coupling. In this way the relation between the gravity coupling constants assumes the standard form 
\begin{equation} 
\label{Newton4e5}
\frac{1}{G_N^{(4)}} = \frac{4\pi}{g  G_N^{(5)}}.
\end{equation}
In addition to the usual KK ansatz, a Scherk-Schwarz twist \cite{Hristov:2014eba, Looyestijn:2010pb} for the gravitino in needed in order to satisfy the BPS conditions in 4d, as noted in \cite{Benini:2020gjh, Hosseini:2017mds}. Thus, to complete the ansatz, we turn on flat gauge connections $\xi^I$ along $y$:
\begin{equation} \label{KK2}
 A_{(5)}^I = A^I_{(4)} + \Re \, z^I (d y - A^0_{(4)}) + \xi^I d y.
\end{equation} 
This twist will bring the extra Killing vector in the 4d reduced theory.\\

Once the reduction is performed one is left with a 4d gauged supergravity with the following salient features (see Appendix \ref{4dt} for a short review of 4d $\mathcal{N}=2$ gauged supergravity).
 The special K\"ahler manifold is specified by the prepotential
  \begin{equation}
    \label{F:4d}
    F(X) = \frac{4}{3\sqrt{6}} C_{IJK} \frac{\check{X}^I \check{X}^J \check{X}^K}{X^0} \qquad \text{with} \qquad \check{X}^I = X^I + \xi^I X^0
  \end{equation}
  and the scalars fields are identified with the special coordinates $z^I = X^I/X^0$. The quaternionic-K\"ahler manifold $\mathcal{QM}$ remains the same as in 5d.\\
  In the 4d theory there are three Killing vectors: two of them are inherited from the 5d theory, while the additional one is given by
  \begin{equation}
    \label{k0:P0}
    K_0^u = \xi^I K^u_I,
  \end{equation}
  which is gauged by the KK vector field $A^0_{(4)}$ in (\ref{KK2}). Similarly, the third Killing prepotential of the theory is given by $\vec{P}_0 = \xi^I \vec{P}_I$.\\
  The 4d electric and magnetic charges are 
  \begin{equation}
    \label{p0:q0}
    \begin{aligned}
      p^0 & = 1, \qquad &  q_0 & = 4G_N^{(4)} g^2 J + \frac{2}{9} C_{IJK} \xi^I \xi^J \xi^K, \\
      p^I & = 0, &  q_{\mathfrak{T}} & = 4 G_N^{(4)} g^2 Q_{\mathfrak{T}} + \frac{1}{3} C_{\mathfrak{T}JK} \xi^J \xi^K,
    \end{aligned}
  \end{equation}
  where the index $\mathfrak{T}$ runs only over the massless vectors $\mathbb{B}^\mathfrak{T}_{\ J} A_\mu^J$ in \eqref{mass:eigen}. The corresponding conserved charges are $Q_{\mathfrak{T}} \equiv Q_J ( \mathbb{B}^{-1})^J_{\ \mathfrak{T}}$.\\
  Electric and magnetic charges form a symplectic vector:
  \begin{equation}
    \label{Q:pq}
    \mathcal{Q} = (p^\Lambda, q_\Lambda), \qquad \Lambda = 0,1,\dots, n_V+1.
  \end{equation}
  Other useful definitions are
  \begin{equation}
    \label{PQ}
    \vec{\mathcal{P}} = (0, \vec{P}_\Lambda), \qquad \vec{\mathcal{Q}} = \langle \vec{\mathcal{P}}, \mathcal{Q} \rangle,
  \end{equation}
  where $\langle V,W \rangle = V_\Lambda W^\Lambda - V^\Lambda W_\Lambda$ is the symplectic-invariant antisymmetric form and the vector are triplets.\\
  Following \cite{Benini:2020gjh}, we impose the ansatz $\vec{\mathcal{Q}} \cdot \vec{\mathcal{Q}} = 1$ and choose a gauge in which
  \begin{equation}
    \label{cond:Q}
    \mathcal{Q}^1 = \mathcal{Q}^2 = 0 \quad \text{and} \quad \mathcal{Q}^3 = -1.
  \end{equation}
  Maxwell's equations at the horizon give the condition
  \begin{equation}
    \label{maxwell:condition}
    \mathcal{K}^u h_{uv} \langle \mathcal{L}^v, \mathcal{Q} \rangle = 0,  
  \end{equation}
  where $\mathcal{K}^u = (0, K^u_\Lambda)$ because we work in a purely electric duality frame.\\
  To make the hyperino variation vanish we have to impose
  \begin{equation}
    \label{hyperino:condition}
    \langle \mathcal{K}^u, \mathcal{V} \rangle = 0,
  \end{equation}
  where $\mathcal{V}(z,\bar{z}) = e^{\mathcal{K}/2} (X^\Lambda, F_\Lambda)$ (see \eqref{sections:4d} and \eqref{hol:sections}) and $F_\Lambda = \frac{\partial{F}}{\partial X^\Lambda}$.\\
  The attractor equations  for the near-horizon limit of 4d BPS static black hole solution are \cite{DallAgata:2010ejj}
  \begin{equation}
    \label{attractor:eqns}
    \frac{\partial}{\partial z^I} \Biggl( \frac{\mathcal{Z}}{\mathcal{L}} \Biggr) = 0 \quad \text{with} \quad \frac{\mathcal{Z}}{\mathcal{L}} = 2i g^2 L_S^2 \quad \& \quad
    \mathcal{Z}= \langle \mathcal{Q}, \mathcal{V} \rangle,  \quad
    \mathcal{L} = \langle \mathcal{P}^3, \mathcal{V} \rangle.
  \end{equation}
  Notice that only $\mathcal{P}^3$ contributes to the superpotential $\mathcal{L}$ after an opportune $SU(2)$ rotation is performed on the prepotentials.\\
The attractor equations are equivalent to
  \begin{equation}
    \label{attractor:eq}
    \partial_\Lambda \biggl[ e^{-\mathcal{K}/2} \biggl( \mathcal{Z}(X) - 2ig^2 L_S^2 \mathcal{L}(X) \biggr) \biggr] = 0,
  \end{equation}
  which is the key formula to reproduce the holographic dual extremization problem that allows to extract the BH entropy from the superconformal index.

\subsection{5d/4d reduction for the LS fixed point }
\label{sub45LS}

We now focus on the model reviewed in section \ref{LSrev}.
  Rewriting the symplectic vector of electric and magnetic charges as 
  \begin{equation}
    \label{Qvec}
    \vec{\mathcal{Q}} = \langle \vec{\mathcal{P}}, \mathcal{Q} \rangle = \vec{\mathcal{P}}_\Lambda \mathcal{Q}^\Lambda - \vec{\mathcal{P}}^\Lambda \mathcal{Q}_\Lambda = \vec{P}_\Lambda \, p^\Lambda,
  \end{equation}
  from \eqref{cond:Q} we obtain the conditions
  \begin{equation}
    \label{BPS1}
    P_0^3 = P_\Lambda^3 \, p^\Lambda = -1, \qquad P^1_0 = P^2_0 = 0,
  \end{equation}
  while Maxwell's equations in \eqref{maxwell:condition} give
  \begin{equation}
    \label{BPS2}
    p^\Lambda K_\Lambda^u = 0 \qquad \Rightarrow \qquad K_0^u = 0.
  \end{equation}
  Solving \eqref{BPS1} and \eqref{BPS2} (using the rotated prepotentials in (\ref{Prot})\footnote{During the computation we are actually using $P_\Lambda^r \to \tilde{P}_\Lambda^r = 2 P_\Lambda^r$ (see \eqref{L6:4d} and the related notes).}) we obtain the following conditions on the fields and the gauge connections:
  \begin{equation}
    \label{sos:flow}
    \sigma = 0, \quad V=1-\theta^2-\tau^2 = 1-\tanh^2\chi, \quad \xi_2 = \frac{\xi_1}{\sqrt{2}}, \quad \xi_1 = -\frac{1}{3\sqrt{2}}.
  \end{equation}
  Finally, from the hyperino variation in \eqref{hyperino:condition}, we have
  \begin{equation}
    \label{sos:hyp}
    \langle \mathcal{K}^u, \mathcal{V} \rangle = \mathcal{K}^u_\Lambda \mathcal{V}^\Lambda - \mathcal{L}^{u\Lambda} \mathcal{V}_\Lambda = e^{\mathcal{K}/2} K_
    \Lambda^u \, X^\Lambda = 0 \qquad \Rightarrow \qquad \sqrt{2} X^2 - X^1 = 0.
  \end{equation}
that  corresponds to the condition imposed on the sections $X^I$  appearing in the massive vector $A^W_\mu$ in \eqref{spectrum:IR}.
 
  Using \eqref{C122}, the prepotential \eqref{F:4d} of our 4d theory becomes
  \begin{equation}
    \label{F4d:our}
    F(X) = \sqrt{2} \, \frac{\check{X}^1 (\check{X}^2)^2}{X^0}, \qquad \text{where} \qquad \check{X}^I = X^I + \xi^I X^0.
  \end{equation}
  Furthermore
  \begin{equation}
    \label{ZL:our}
    \begin{split}
      e^{-\mathcal{K}/2} \mathcal{Z}(X) & = q_\Lambda \, X^\Lambda -F_0 = \hat{q}_\Lambda \, X^\Lambda + \sqrt{2} \, \frac{X^1 (X^2)^2}{(X^0)^2}\\
      e^{-\mathcal{K}/2} \mathcal{L}(X) & = P_\Lambda^3 \, X^\Lambda = - X^0 + \frac{3}{\sqrt{2}} \, (2-\cosh^2\chi) \, X^1 + 3  \cosh^2 \chi \, X^2,    
      \end{split}
    \end{equation}
    where
    \begin{equation}
      \label{qhat}
      \begin{split}
        \hat{q}_I & = q_I - \frac{1}{3} C_{IJK} \xi^J \xi^K, \\
        \hat{q}_0 & = q_0 - \frac{2}{9} C_{IJK} \xi^I \xi^J \xi^K
      \end{split}
    \end{equation}
    and in the last line we have introduced the new coordinate
    $\chi$ defined in (\ref{paramflow}).
    Thus, from \eqref{attractor:eq} we obtain the following set of equations:
    \begin{equation}
      \label{attractor:eqns}
      \begin{split}
        \partial_\Lambda & \Biggl[ \sqrt{2} \, \frac{X^1 (X^2)^2}{(X^0)^2} + \hat{q}_\Lambda \, X^\Lambda \!-\! 2i g^2 L_s^2 \biggl( 3\sqrt{2} \, X^1 \!-\! X^0\!-\! 3 \cosh^2 \chi \Bigl( \frac{X^1}{\sqrt{2}} \!-\!  X^2 \Bigr) \!-\!\alpha \biggr) \Biggr] \!\!= \!0,\\
          \frac{\partial}{\partial L_s^2} & \Biggl[ \sqrt{2} \, \frac{X^1 (X^2)^2}{(X^0)^2} + \hat{q}_\Lambda \, X^\Lambda  \!-\! 2i g^2 L_s^2 \biggl(   3\sqrt{2} \, X^1 \!-\!X^0 - 3 \cosh^2 \chi \Bigl( \frac{X^1}{\sqrt{2}} \!-\!  X^2 \Bigr) \!-\!\alpha \biggr) \Biggr] \!\!=\! 0.
      \end{split}
    \end{equation}
    In the first line we rewrote \eqref{attractor:eq}, adding the constant $\alpha$ that does not modify the equations, while the second line fixes the gauge $\mathcal{L}=\alpha$.\\
    The variation in \eqref{attractor:eqns} with respect to $X^2$ gives the equation
    \begin{equation}
      \label{var:X2}
      2\sqrt{2} \, \frac{X^1 X^2}{(X^0)^2} + \hat{q}_2 -6 i g^2 L_s^2 \cosh^2\chi = 0,
    \end{equation}
    that determines $\chi$ and $q_2$ in terms of the sections and of $L_s$.\\
    We can now use the hyperino condition \eqref{sos:hyp} to eliminate $X^2$ from the equations in \eqref{attractor:eqns}. The remaining equations are equivalent to the conditions of extremization of the function
    \begin{equation}
      \label{S:1}
      \mathcal{S} = \beta \Biggl[ \frac{\sqrt{2}}{2} \frac{(X^1)^3}{(X^0)^2} + \hat{q}_0 X^0 - \sqrt{3} \hat{q}_R X^1 - 2i g^2 L_s^2 \biggl( 3\sqrt{2} X^1- X^0 - \alpha \biggr) \Biggr]
    \end{equation}
    w.r.t. the variables $X^0,X^1$ and $L_s$. $\beta$ is a constant that will be useful later and $\hat{q}_R$ is the charge with respect to the massless vector $A_R$ in \eqref{mass:eigen}:
    \begin{equation}
      \label{qR}
      \hat{q}_R = \frac{1}{\sqrt{6}} \bigl( \sqrt{2} \hat{q}_1 + \hat{q}_2 \bigr) = 4 g^2 G_N^{(4)} Q_R.
    \end{equation}
    Notice that we have divided by $2^{2/3}$ the coefficients of $K_R$, in order to have a proper normalization of the AdS scale. In fact, the relation between $L$ and the coupling constant $g$ can be red from the value of the scalar potential at the IR critical point $V|_{\text{IR}} = -6 \cdot 2^{4/3}$, from which
\begin{equation}
  \label{L:g}
  L = \frac{1}{2^{2/3}g}.
\end{equation}
    
    $\mathcal{S}$ is an homogeneous function of degree 1 in $X^\Lambda$, except for the term involving $\alpha$. Therefore
    \begin{equation}
      \label{S:crit}
      \mathcal{S} \big|_{\text{crit}} = 2i\alpha\beta g^2L_s^2
    \end{equation}
    at the critical point and choosing
    \begin{equation}
      \label{alphabeta}
      \alpha\beta = \frac{\pi}{2i G_N^{(4)} g^2},
    \end{equation}
    we obtain that $\mathcal{S} \big|_{\text{crit}}$ corresponds to the black hole entropy.
    Using \eqref{p0:q0} and \eqref{qhat}, the extremization problem \eqref{S:1} becomes
 \begin{equation}
  \label{S:2}
        \mathcal{S} = \frac{1}{\alpha} \Biggl[ \frac{\pi}{2 i g^2 G_N^{(4)}} \frac{\sqrt{2}}{2} \frac{(X^1)^3}{(X^0)^2} - 2\pi i \biggl( J  X^0 + \sqrt{3} Q_R  X^1 \biggr) - 2\pi i \Lambda \biggl( 3\sqrt{2} X^1 - X^0 - \alpha \biggr) \Biggr],
      \end{equation}
       where we have redefined the Lagrange multiplier $L_s^2 = 2i G_N^{(4)} \Lambda$.\\

\section{Field theory results and holographic matching}
\label{FTHOLM}
In this section we study the extremization problem from the dual field theory perspective.
The entropy function derived above is extracted from the superconformal index, i.e.
the supersymmetric partition function on $S^3 \times S^1$.
It is  used to compute the entropy function of the  rotating black hole, once complex fugacities are considered.
The two most popular approaches to this calculations correspond to the Cardy-like limit from the saddle point 
approximation of the matrix integral developed in \cite{Choi:2018hmj} and to the expansion of the index in terms of the solutions of the BAEs developed in \cite{Benini:2018ywd}.
In the first case an opportune limit of large conserved charges is considered.
This limit has been shown to reproduce the entropy function, that after a  Legendre transform becomes the 
entropy of the dual black hole. 
The complementary approach, discussed in \cite{Benini:2018ywd} consists of computing the large $N$ index (see \cite{Benini:2021ano,Lezcano:2021qbj} for recent discussions and results at finite $N$) with complex fugacities without considering any  limits on the conserved charges.
While the Cardy-like limit treats the index as a matrix integral and the entropy function is extracted from a saddle point approximation, in this second case the index is written as a sum of residues, where the poles are defined by solutions of 
the BAEs. 
The two approaches have in general different regimes of validity but they have been shown to agree and to reproduce the BH entropy.
The analysis of the Cardy like limit of the SCI for the LS fixed point was worked out in \cite{Amariti:2020jyx}. Here we 
will give a different derivation of the result,  showing how to extract the leading contribution to the entropy function from the BAEs. Then  
we compare the results against  the one obtained from the supergravity calculation, by exploiting the holographic dictionary.

\subsection{Entropy function for the LS fixed point from the  BAE}
\label{subft}
The superconformal index corresponds to the Witten index calculated on a three sphere and it counts states that are
annihilated by the supersymmetry generators on the curved space.
In the case of the LS fixed point the index can be formally expressed as the sum
\begin{equation}
\mathcal{I} =Tr (-1)^F \, p^{J_1+\frac{1}{2} (R_1+R_2)} q^{J_2+\frac{1}{2} (R_1+R_2)} v^{R_1-R_2},
 \end{equation}
where $J_1$ and $J_2$ are two angular momenta on the three sphere, and $R_{1,2}$ are two R-charge generators assigning 
$R$-charge two to $\Phi_{1,2}$ and zero to $\Phi_{2,1}$ respectively. The exact $R$-symmetry is then $\frac{R_1+R_2}{2}$
while $R_1-R_2$ is a flavor symmetry.
The fermionic number can be equivalently written as $(-1)^F = (-1)^{\frac{R_1+R_2}{2}}$.
The fugacities $p,q,v$ can be expressed in terms of the chemical potentials $\tau,\sigma,\xi$ by the relations
\begin{equation}
p=e^{2 \pi i \tau}, \quad q=e^{2 \pi i \sigma}, \quad v=e^{2 \pi i \xi}
 \end{equation}
and by defining 
$
y_I=e^{2 \pi i \Delta_I}$ with $\Delta_1=\xi_1 + \frac{\sigma+\tau}{4}
$
the index can be written as 
\begin{equation}
Tr_{BPS}  \, p^{J_1} q^{J_1} y_1^{R_1} y_2^{R_2}
\end{equation}
with the constraint $\tau+\sigma-2\Delta_1-2\Delta_2  \in 2 \mathbb{Z}+1$.
In order to match the holographic results in the rest of the discussion we further restrict to the case $\sigma=\tau \equiv \omega$.
The index can be computed as a sum over the solutions
to a set of BAE, and it assumes the standard  form  \cite{Closset:2017bse,Benini:2018mlo}
\begin{equation}
\mathcal{I} = \kappa_{N_c} \sum_{\hat u} \mathcal{Z}(\hat u; \Delta,\omega) H(\hat u; \Delta,\omega)
\end{equation}
where $\hat u$ represents a set of solutions of the BAE and 
\begin{eqnarray}
\kappa_{N_c} &=& \frac{1}{N_c!} (q,q)_{\infty}^{2 N_c} \prod_{I=1}^{2} \tilde \Gamma (\Delta_I; \omega)^{N_c}\, ,
\nonumber \\
H(\hat u;\Delta,\omega) &=& \det \left( \frac{1}{2 \pi i } \frac{\partial (Q_1,\dots,Q_{N_c})}{\partial(u_1\dots,u_{N_c-1},\lambda)}\right)\, ,
\nonumber \\
\mathcal{Z}  &=& \prod_{i \neq j} \frac{ \prod_{a=1}^{2} \tilde \Gamma(u_{ij}+ \Delta_a,\omega)}{\tilde \Gamma(u_{ij},\omega)}
\, , 
\nonumber \\
Q_i(u;\Delta,\omega) &=& e^{2 \pi i (\lambda+2\sum_j u_{ij})} \prod_{j=1}^{N} \frac{\theta_0(u_{ji}+\Delta_a;\omega)}{\theta_0(u_{ij}+\Delta_a;\omega)}\, .
 \end{eqnarray}
 In the formulas above we defined the difference between two holonomies as $u_{ij} = u_i - u_j$.
The modified elliptic Gamma functions $\tilde \Gamma$ are related to the standard elliptic Gamma functions $\Gamma_e$ by
\begin{equation}
\tilde \Gamma(u;\tau,\sigma) 
\equiv
\tilde \Gamma(u) 
=
 \Gamma_e(e^{2 \pi i u};e^{2 \pi i \tau},e^{2 \pi i \sigma})\ ,
\end{equation}
and
\begin{equation}
\Gamma_e(z;p,q) \equiv \prod_{j,k=0}^\infty 
\frac{1-p^{j+1}q^{k+1} /z}{1-p^j q^k z} \equiv \Gamma_e(z) \ .
\end{equation}
The elliptic theta function  $\theta_0 (u;\omega)= (e^{2 \pi i u};e^{2 \pi i \omega})_\infty (e^{2 \pi i (\omega-u)};e^{2 \pi i \omega})_\infty $ is expressed in terms of the Pochhammer symbols $(a,b)_\infty = \prod_{k=0}^{\infty} (1-a b^k)$.

The index is evaluated on the solutions $\hat u$ of the BAE given by
 \begin{equation}
1 = e^{2 \pi i \lambda} 
\prod_{i (\neq j)=1}^{N} 
\frac
{\theta_0(u_{ij} + \Delta_1;\omega) \theta_0(u_{ij} + \Delta_2;\omega)}
{\theta_0(u_{ji} + \Delta_1;\omega) \theta_0(u_{ji} + \Delta_2;\omega)}.
 \end{equation}
The discrete solutions for generic values of the charges $\Delta_I$ are the same found for  the $\mathcal{N}=4$ SYM  and they have been originally discussed in \cite{Hosseini:2016cyf,Hong:2018viz}:
\begin{equation}
u_{ij}  = \frac{\tau }{N_c} (j-i), \quad  u_j = \frac{\tau(N_c-j)}{N_c}+ \overline u, \quad \lambda =\frac{N_c-1}{2}
 \end{equation}
with the constant $\overline u$ enforcing the $SU(N)$ constraint. This is consistent with the fact that the discrete solutions
for the $\mathcal{N}=4$ index are independent from the value of the charges $\Delta_I$. 
Plugging the solution in $\mathcal{Z}$, the leading order of the index  $\mathcal{I}$ is
\begin{equation}
\log \mathcal{I} =
-\frac{i \pi  N_c^2  \{\Delta _1\} 
\{\Delta _2\} (\{\Delta _1 \}+\{\Delta _2\}
)}{\omega ^2}
\end{equation}
where the $\tau$-modded charges $\{\Delta_I\}$ 
are constrained by
\begin{equation}
\label{constr:delta}
2\{ \Delta_1 \} + 2\{\Delta_2 \} -2\omega = \pm 1. 
\end{equation}

Starting from here we can also set the extremization problem that we need to match with the supergravity
results. The Legendre transform of the logarithm of the superconformal index, referred to as the entropy function, gives rise to the black hole entropy of the dual gravitational theory.

In order to set such entropy function and the extremization problem we then need to add to $\log {\mathcal{I}}$
the conjugate variables and the constraint (\ref{constr:delta}).
The conjugate variables are referred to as $r_{1,2}$ for the charges $\{ \Delta_{1,2} \} $ and  $J$ for $\omega$.
All in all the extremization problem corresponds to the entropy
\begin{eqnarray}
 \mathcal{S}&=& -i \pi N_c^2 \frac{ \{\Delta _1\} 
\{\Delta _2\} (\{\Delta _1 \}+\{\Delta _2\})}{\omega^2} \nonumber \\
&+& 2 \pi i (2 J \omega + 2 r_1 \{\Delta _1\} +2 r_2 \{\Delta _2\} ) 
 \nonumber \\
 &-&2 \pi i \Lambda (2\{\Delta _1\} +2\{\Delta _2\}   -2\omega \pm 1)
\end{eqnarray}
We then further  restrict the charges by imposing $\{ \Delta_1 \}  = \{\Delta_2 \} \equiv X_r $.
The conjugate variable for the charge $X_r$ is now referred to as $r$ (obtained from $r_1=r_2\equiv r$) and the entropy function becomes
\begin{equation}
\label{SEfin}
 \mathcal{S}= -2i \pi N_c^2 \frac{X_r^3}{\omega^2} + 2 \pi i (2 J \omega + 4 r X_r) -2 \pi i \Lambda (4 X_r-2\omega \pm 1)
\end{equation}
where $\Lambda$ is a Lagrange multiplier enforcing the constraint  (\ref{constr:delta}).

From here the evaluation of the black hole entropy is straightforward and matches with the one obtained from the Cardy like limit of the SCI 
in \cite{Amariti:2020jyx}.
Similar considerations can be done for $\sigma\neq \tau$ and beyond the leading order in the expansion, even if we are interested in just this contribution, because we match with supergravity at lowest order.

\subsection{Holographic matching}

The last step consists in pointing out the AdS/CFT dictionary between the charges in gravity and field theory  and check the agreement between the results in sub-section \ref{sub45LS} and the ones in \ref{subft}. First, the number of colors $N_c$ of the gauge group in field theory is related to the Newton constant $G_N^{(5)}$ . In fact
\begin{equation}
  \label{G5:N2}
  G_N^{(5)} =  \frac{L^3 \text{Vol}(Y_5)}{2\pi^2 N_c^2}, \qquad \text{with} \qquad \text{Vol}(Y_5) = \frac{\pi^3 N_c^2}{4a},
\end{equation}
where $L$ is the AdS$_5$ length scale. 
For the 4d SCFT LS fixed point
\begin{equation}
  \label{a:N}
  a = \frac{27}{128}N_c^2.
\end{equation}
Using \eqref{L:g} and \eqref{a:N} we obtain
\begin{equation}
      \label{G5}
     G_N^{(5)} = \frac{4 \pi}{27 g^3 N_c^2}
   \end{equation}
and, using \eqref{Newton4e5}, we can relate the Newton constant in 4d to $N_c$:
    \begin{equation}
      \label{G4}
      G_N^{(4)} =  \frac{1}{27 g^2 N_c^2}.
    \end{equation}
    The angular momentum $J$ on the gravity side corresponds to the one on the field theory side, while 
     the electric charge $r$ is related to $Q_R$ by
    \begin{equation}
      \label{QR:r}
      r = \frac{Q_R}{\gamma}.
    \end{equation}
    The coefficient $\gamma$ can be inferred by comparing the 't Hooft anomalies for global currents in the boundary field theory with the Chern-Simons couplings for gauge fields in the bulk 5d gravity:
    \begin{equation}
      \label{anomalies}
      \frac{g^3}{24\pi^2} \Tr (Q_I Q_J Q_K) = \frac{1}{8 \pi G_N^{(5)}} \frac{2}{3\sqrt{6}} \cdot \frac{1}{3} C_{IJK}.
    \end{equation}
    Inserting $C_{RRR}=6$, we obtain $\gamma= \sqrt{6}$.
    Finally, after the change of coordinates $X^0 \to 2 \alpha \omega$ and $X^1 \to \frac{2\sqrt{2}}{3} \, \alpha  X_r $, our entropy  exactly matches the one obtained from the large $N_c$ limit of the superconformal index of the dual theory
    given in formula (\ref{SEfin}).
    
\section{Conclusions}
\label{concl}

In this paper we have studied  the entropy function for a conjectured BPS Kerr-Newman black hole
originating from the 5d truncation of the LS fixed point proposed in \cite{Ceresole:2001wi}.
Our analysis is along the lines of \cite{Benini:2020gjh}, that generalized the results of \cite{Hosseini:2017mds} in presence of hypermultiplets.
After an opportune local rotation on the Killing prepotential and a generalization of the KK ansatz we have obtained the entropy function and we have matched such results from the one expected on the field theory side \cite{Amariti:2020jyx}. 
We further computed such an entropy function from the SCI using the formalism of \cite{Benini:2018ywd}, by solving the  BAE.

Let us conclude with some open questions and speculations.
It should be interesting to apply the procedure of \cite{Benini:2020gjh} to other consistent  truncation 
with a dual interpretation in order to validate the results obtained from the field theory side.
For example the truncation of the LS fixed point discussed in  \cite{Bobev:2014jva} is very interesting because 
it involves two massless vectors, i.e. the full global symmetry is visibile in such a truncation.
Such truncation has however a non abelian gauging and a further truncation is needed to get an abelian gauging in order to study it in the formalism of 
\cite{Benini:2020gjh}. It should be interesting to recast such a truncation in the language of \cite{Ceresole:2000jd} and then apply the procedure discussed here in order to obtain an entropy function from supergravity matching the one with the global symmetry  turned on .
It would be also interesting to investigate on the possible interpretation of the flow relating the UV and the IR fixed points 
studied in \cite{Ceresole:2001wi} after the KK reduction (see also related ideas in \cite{GonzalezLezcano:2022mcd}, where a field theoretical interpretation of the flow across dimensions has been proposed).
Another interesting open question regards the analysis of higher derivatives corrections, in order to reproduce 
similar corrections depending on the gravitational anomaly, obtained from the field theory side (see \cite{Cassani:2022lrk,Bobev:2022bjm}).

\section*{Acknowledgments}
We are grateful to Marco Fazzi and Federico Faedo for comments and discussions.
This work has been supported in part by the Italian Ministero dell'Istruzione, 
Universit\`a e Ricerca (MIUR), in part by Istituto Nazionale di Fisica Nucleare (INFN) through the “Gauge Theories, Strings, Supergravity” (GSS) research project and in part by MIUR-PRIN contract 2017CC72MK-003.

\appendix

\section{5d $\mathcal{N}=2$ Abelian gauged supergravity}\label{5dt}

We briefly review the general form of $d=5$ $\mathcal{N}=2$ gauged supergravity coupled to $n_V$ vector multiplets and $n_H$ hypermultiplets \cite{Ceresole:2001wi, Gunaydin:1983bi, Ceresole:2000jd, Lauria:2020rhc}. Each vector multiplet contains a vector $A_\mu^x$ $(x=1, \dots, n_V)$, a gaugino and a scalar $\phi^x$, each hypermultiplet contains a hyperino and four real scalars $q^u$ $(u = 1, \dots, 4n_H)$, and finally the graviton multiplet contains a graviton, a gravitino $\psi_{\mu i}$ $(i=1,2)$ and a vector $A_\mu^0$. Therefore we indicate the gauge fields of the theory with $A_\mu^I, \ I = 1, \dots, n_V+1$.\\

The scalars of the vector multiplets and of the hypermultiplets parametrize a manifold $\mathcal{M}$, which is the direct product of a very special real manifold $\mathcal{SM}$ and a quaternionic-K\"ahler manifold $\mathcal{QM}$
\begin{equation}
  \label{Mscalar}
  \mathcal{M} = \mathcal{SM} \otimes \mathcal{QM},
\end{equation}
with dim$_{\mathbb{R}} \mathcal{SM} = n_V$ and dim$_{\mathbb{R}} \mathcal{QM} = 4 n_H$.\\
The last geometrical data needed to define the theory is the gauging, achieved by identifying the gauge group $K$ as a subgroup of the isometries of $\mathcal{M}$. In particular, we will focus on the case in which $K=U(1)^{n_V+1}$ and $\mathcal{QM}$ has $n_V+1$ Abelian isometries that are gauged by introducing $n_V+1$ Killing vectors.\\

The bosonic part of the Lagrangian is
\begin{align}
  \label{L:5d}
    e^{-1} \mathscr{L} = &  \frac{R}{2} -\frac{1}{2} \, g_{xy}(\phi) \, \partial_\mu \phi^x \, \partial^\mu \phi^y  - \frac{1}{2} h_{uv}(q) \mathcal{D}_\mu q^u \, \mathcal{D}^\mu q^v - \frac{1}{4} a_{IJ}(\phi) F^I_{\mu\nu} F^{J \, \mu\nu} \notag \\
    & + \frac{e^{-1}}{6 \sqrt{6}} C_{IJK} \epsilon^{\mu\nu\rho\sigma\tau} F_{\mu\nu}^I F_{\rho\sigma}^J A_\tau^K - g^2 V(\phi,q),
\end{align}
where $R$ is the scalar curvature, $F^I_{\mu\nu} = \partial_\mu A_\nu^I - \partial_\nu A_\mu^I$ is the Abelian field strength, $g$ is a coupling constant and $V$ is the scalar potential. We will explain the other terms in the following paragraphs.

\subsection{Very special real manifold}
It can be described as the submanifold
\begin{equation}
  \label{SM}
  \mathcal{V}(h) \equiv \Bigl\{ C_{IJK} h^I(\phi) h^J(\phi) h^K(\phi) =1 \Bigr\} \subset \mathbb{R}^{n_V+1}
\end{equation}
specified by the totally symmetric constant tensor $C_{IJK}$ that determines also the Chern-Simons couplings of the vector fields. $\phi^x$ are real coordinates on $\mathcal{SM}$ and $h^I$ are coordinates on the ambient space $\mathbb{R}^{n_V+1}$ that give rise to sections $h^I(\phi^x)$ on the very special manifold.\\
The coefficient $C_{IJK}$ also determine the metrics for vector fields and vector multiplet scalar fields:
\begin{equation}
  \label{aIJ}
  a_{IJ} \equiv  -2 \, C_{IJK} k^K + 3 \, C_{IKL} C_{JMN} k^K h^L h^M h^N = h_I h_J + h_{xI} h^x_J,
\end{equation}
\begin{equation}
  \label{gxy:aIJ}
  g_{xy} \equiv h^I_x h^J_y a_{IJ},
\end{equation}
where
\begin{equation}
  \label{hIx}
  h^I_x \equiv - \sqrt{\frac{3}{2}} \partial_x h^I(\phi).
\end{equation}
We also report the useful relations
\begin{equation}
  \label{hdh}
  h^I h_I = 1, \qquad h^I_x h_I = h_{Ix}h^I = 0.
\end{equation}

\subsection{Quaternionic-K\"ahler manifold}
The quaternionic-K\"ahler geometry is determined by $4n_H$-beins $f^{iA}_u$, where $i=1,2$ is an $SU(2)$ index and $A=1,\dots, 2n_H$ is an $USp(2n_H)$ index. The metric is given by
\begin{equation}
  \label{huv}
  h_{uv} \equiv f^{\ iA}_u f^{\ jB}_v \varepsilon_{ij} C_{AB} = f^{\ iA}_u f_{v \, iA},
\end{equation}
where $\varepsilon_{ij}$ and $C_{AB}$ are the invariant tensors of $SU(2)$ and $USp(2n_H)$ that raise and lower the indices. The vielbeins are covariantly constant:
\begin{equation}
  \label{var:f}
  0 = \partial_uf_v^{\ iA} - \Gamma_{uv}^{\ \ w} f_w^{\ iA} + f_v^{\ iB} \omega_{uB}^{\ \ A} + \omega_{uk}^{\ \ i} f_v^{\ kA},
\end{equation}
where $\Gamma_{uw}^{\ \ v}$ is the Levi-Civita connection, $\omega_{u \ A}^{\ B}$ is the $USp(2n_H)$ connection and $\omega_{ui}^{\ \ j}$ is the $SU(2)$ connection.\\
The scalars $q^u$ are real coordinates on $\mathcal{QM}$.\\
 
The $SU(2)$ curvature is given by
\begin{equation}
  \label{R:SU2}
  \mathcal{R}_{uv \, ij} = f_{u \, C(i}f_{j)v}^{\ C}.
\end{equation}
These curvatures are proportional to the three complex structures of the quaternionic space, and thus they satisfy the quaternionic relation
\begin{equation}
  \label{RR}
  (\mathcal{R}^r)_{uv} (\mathcal{R}^s)^{vw} = -\frac{1}{4} \delta^{rs} \delta_u^{\ w} - \frac{1}{2} \varepsilon^{rst}  (\mathcal{R}^t)^{\ w}_u,
\end{equation}
with $r,s,t=1,2,3$ and $\mathcal{R}^r = \frac{i}{2} \mathcal{R}_{ij} ( \sigma^r)^{ij} $\footnote{$(\sigma^r)_i^{\ j}$ are the usual $\sigma$ matrices and $(\sigma^r)^{ij}$ are defined using the NW-SE contraction convention.}.

\subsection{Gauging}
We only consider gaugings of Abelian isometries of $\mathcal{QM}$ by the vectors $A^I_\mu$. These isometries are generated by the Killing vectors $k_I^u(q)$, which can be determined in terms of the Killing prepotentials $P_I^r(q)$, satisfying\footnote{\label{P2P}Notice that, due to the relations \eqref{RR} and \eqref{k:P}, our Killing prepotentials are defined with an extra $\frac{1}{2}$ w.r.t. the definition used in \cite{Benini:2020gjh}, i.e. $\vec{P}_I^{\text{ our}} = \frac{1}{2} \vec{P}_I^{\text{ BCSZZ}}$.}: 
\begin{equation}
  \label{k:P}
  (\mathcal{R}^r)_{uv} k^v_I = \nabla_u P_I^r, \qquad \nabla_u P^r_I \equiv \partial_u P_I^r + 2 \, \varepsilon_{rst} \, \omega_u^s P_I^t.
\end{equation}
These prepotentials also obey the constraint
\begin{equation}
  \label{RKK:PP}
  \frac{1}{2}( \mathcal{R}^r)_{uv} k_I^u k_J^v - \varepsilon^{rst} P_I^s P_J^t + \frac{1}{2} f_{IJ}^K P_K^r = 0,
\end{equation}
where $f_{IJ}^K$ are the gauge group structure constants.\\

The covariant derivative in \eqref{L:5d} is defined as
\begin{equation}
  \label{cov:q}
  \mathcal{D}_\mu q^u = \partial_\mu q^u + g \, A_\mu^I k_I^u.
\end{equation}

Finally the scalar potential is given by
\begin{equation}
  \label{V:5d}
    V  = -4 \, P_I^r P_J^r h^I h^J + 3 \, P_I^r P_J^r \, g^{xy} \partial_x h^I \partial_y h^J + \frac{3}{4} h_{uv} k_I^u k_J^v h^I h^J.
  \end{equation}

\section{4d $\mathcal{N}=2$ Abelian gauged supergravity}\label{4dt}

We outline the main features of $d=4 \ \mathcal{N}=2$ gauged supergravity coupled to $n_V$ vector multiplets and $n_H$ hypermultiplets \cite{Benini:2020gjh, Andrianopoli:1996cm, Andrianopoli:1996vr, Lauria:2020rhc}. Each vector multiplet contains a vector $A_\mu^i \ (i=1, \dots, n_V)$, a doublet of gauginos and a complex scalar $z^i$, each hypermultiplet contains a doublet of hyperinos and for real scalars $q^u \ (u=1, \dots, 4n_H)$, and the graviton multiplet contains a graviton, a doublet of gravitinos and a vector $A_\mu^0$. We indicate the gauge fields of the theory with $A_\mu^\Lambda, \ \Lambda = 0, \dots, n_V$.\\

The scalars of the vector multiplets and of the hypermultiplets parametrize a manifold $\mathcal{M}$, which is a direct product of a special K\"ahler manifold $\mathcal{KM}$ and a quaternionic-K\"ahler manifold $\mathcal{QM}$
\begin{equation}
  \label{Msc:4d}
  \mathcal{M} = \mathcal{KM} \otimes \mathcal{QM},
\end{equation}
with dim$_{\mathbb{C}}\mathcal{KM} = n_V$ and dim$_{\mathbb{R}} \mathcal{QM} = 4 n_H$.\\
The gauging is achieved by identifying the gauge group $K$ as a subgroup of the isometries of $\mathcal{M}$. As in the 5d theory, we focus on the case in which $K=U(1)^{n_V+1}$ and $\mathcal{QM}$ has $n_V+1$ Abelian isometries gauged by $n_V+1$ Killing vectors.\\

It is always possible to choose a duality frame in which all gaugings are purely electric. In these frames the bosonic part of the Lagrangian becomes
\begin{equation}
  \label{L:4d}
  \begin{split}
    e^{-1} \mathscr{L} & = \frac{R}{2} - G_{i \bar{\jmath}} (z, \bar{z}) \, \partial_\mu z^i \, \partial^\mu z^{\bar{\jmath}} - \frac{1}{2} h_{uv}(q) \, \mathcal{D}_\mu q^u \, \mathcal{D}^\mu q^v \\
    & + \frac{1}{8} \Im \, \mathcal{N}_{\Lambda\Sigma} F^\Lambda_{\mu\nu} F^{\Sigma \, \mu\nu} - \frac{e^{-1}}{16} \Re \, \mathcal{N}_{\Lambda\Sigma} (z, \bar{z}) F^\Lambda_{\mu\nu} F^\Sigma_{\rho\sigma} \epsilon^{\mu\nu\rho\sigma} - g^2 V(z,\bar{z},q).
  \end{split}
\end{equation}
We are mostly using the same notation as in Appendix \ref{5dt}.

\subsection{Special K\"ahler manifold}
The complex scalar fields $z^i$ are coordinates on $\mathcal{KM}$, that is a K\"ahler-Hodge manifold. This is a K\"ahler manifold with a K\"ahler potential $\mathcal{K}$ and metric
\begin{equation}
  \label{Kahler:metric}
  G_{i\bar{\jmath}} = \partial_i \partial_{\bar{\jmath}} \mathcal{K} (z,\bar{z}).
\end{equation}
Furthermore, it has the property that there is a line bundle $\mathcal{L}$ whose first Chern class coincides with the K\"ahler class of the manifold.\\
An alternative definition of a special K\"ahler manifold can be obtained by constructing a flat $2n_V+2$-dimensional symplectic bundle over the K\"ahler-Hodge manifold whose generic sections
\begin{equation}
  \label{sections:4d}
  \mathcal{V} =
  \begin{pmatrix}
    L^\Lambda\\
    M_\Lambda
  \end{pmatrix},
  \qquad \Lambda = 0, \dots, n_V,
\end{equation}
are covariantly holomorphic
\begin{equation}
  \label{cov:hol}
  D_{\bar{\imath}} \mathcal{V} \equiv \partial_{\bar{\imath}} \mathcal{V} - \frac{1}{2} \bigl( \partial_{\bar{\imath}} \mathcal{K} \bigr) \mathcal{V} = 0
\end{equation}
and obey the further constraints
\begin{equation}
  \label{constr1:4d}
  i \langle \mathcal{V}, \bar{\mathcal{V}} \rangle \equiv i \bigl( M_\Lambda \bar{L}^\Lambda - L^\Lambda \bar{M}_\Lambda \bigr)  = 1
\end{equation}
and
\begin{equation}
  \label{constr2:4d}
  \langle \mathcal{V}, D_i \mathcal{V} \rangle = 0,
\end{equation}
where $i\langle \ , \ \rangle$ denotes the $Sp-$invariant antisymmetric form. The K\"ahler potential can be computed as a symplectic invariant: introducing the holomorphic sections
\begin{equation}
  \label{hol:sections}
    \Omega  = e^{-\mathcal{K}/2} \mathcal{V} = e^{-\mathcal{K}/2}
    \begin{pmatrix}
      L^\Lambda\\
      M_\Lambda
    \end{pmatrix}
    \equiv
    \begin{pmatrix}
      X^\Lambda \\
      F_\Lambda
    \end{pmatrix}
    \quad \text{such that} \quad \partial_{\bar{\imath}} \Omega  = 0,
  \end{equation}
  the relation \eqref{sections:4d} gives
\begin{equation}
  \label{Kahler:pot}
  \mathcal{K} = - \log \bigl[ i \bigl( \bar{X}^\Lambda F_\Lambda - X^\Lambda \bar{F}_\Lambda \bigr) \bigr].
\end{equation}
Notice that under K\"ahler transformations:
\begin{equation}
  \label{Kahler:transf}
  \mathcal{K} \to \mathcal{K} + f + \bar{f} \quad \text{and} \quad \Omega \to \Omega e^{-f}.
\end{equation}
Thus, since $X^\Lambda \to X^\Lambda e^{-f}$, if the matrix $e^\lambda_{\ i}(z) = \partial_i (X^\lambda/X^0)$ (with $\lambda = 1, \dots, n_V$) is invertible, we can use $X^\Lambda$ as homogeneous coordinates on $\mathcal{KM}$, at least locally. In this case, we can set
\begin{equation}
  \label{F:FX}
  F_\Lambda = F_\Lambda(X)
\end{equation}
and the constraint \eqref{constr2:4d} implies the integrability condition
\begin{equation}
  \label{int:cond}
  \frac{\partial F_\Sigma}{\partial X^\Lambda} - \frac{\partial F_\Lambda}{\partial X^\Sigma} = 0.
\end{equation}
Therefore, the sections $F_\Lambda$ are the derivatives of a holomorphic homogeneous function $F(X)$ of degree 2 called prepotential, i.e.:
\begin{equation}
  \label{FL}
  F_\Lambda = \frac{\partial F(X)}{\partial X^\Lambda}.
\end{equation}
In these frames the geometry is completely specified by the prepotential and the coordinates $z^i \equiv X^i/X^0$, with $i=1,\dots,n_V$, are called special coordinates.\\

The couplings between the scalars $z^i$ and the vector fields are determined by the period matrix $\mathcal{N}$, defined via the relations
\begin{equation}
  \label{N:rel}
  M_\Lambda = \mathcal{N}_{\Lambda\Sigma} L^\Sigma, \qquad D_{\bar{\imath}} \bar{M}_\Lambda = \mathcal{N}_{\Lambda\Sigma} D_{\bar{\imath}} \bar{L}^\Sigma.
\end{equation}
When a prepotential exists, the period matrix is obtained from
\begin{equation}
  \label{periodmatrix}
  \mathcal{N}_{\Lambda\Sigma} = \bar{F}_{\Lambda\Sigma} +2i \frac{(\Im F_{\Lambda\Gamma}) X^\Gamma (\Im F_{\Sigma\Delta}) X^\Delta}{X^\Psi (\Im F_{\Psi\Omega})X^\Omega},
\end{equation}
where $F_{\Lambda\Sigma} = \partial_\Lambda \partial_\Sigma F$.

\subsection{Quaternionic-K\"ahler manifold and gauging}
The quaternionic-K\"ahler manifold has been already summarized in Appendix \ref{5dt}.\\
As before, we focus on the case in which $\mathcal{QM}$ has $n_V+1$ Abelian isometries gauged by the same number of Killing vectors. In general, one could have both electric and magnetic gaugings, described by Killing vectors $k^u_\Lambda$ and $k^{u \Lambda}$ that transform as a vector under $Sp(n_V+1, \mathbb{R})$ duality transformations. Nevertheless, one can always move to a duality frame in which all gaugings are purely electric. We will work in such a frame.\\

The scalar potential is given by
\begin{equation}
  \label{scalar:pot:4d}
    V  = - P_\Lambda^r P_\Sigma^r \Bigl( \bigl(\Im \, \mathcal{N} \bigr)^{-1 | \Lambda \Sigma} + 8 \, e^{\mathcal{K}} X^\Lambda \bar{X}^\Sigma \Bigr)  + 4 \, e^{\mathcal{K}} h_{uv} k_\Lambda^u k_\Sigma^v X^\Lambda \bar{X}^\Sigma.
\end{equation}

\section{Revisiting KK reduction}\label{KK}

In this appendix we revisit the KK reduction from the 5d Lagrangian \eqref{L:5d} 
to the 4d one \eqref{L:4d}. Our analysis generalizes the one performed in the 
Appendix D of \cite{Benini:2020gjh}, giving a recipe that allows to construct a 4d Lagrangian in the formalism of \cite{Benini:2020gjh, Hosseini:2017mds} starting from a 5d Lagrangian written in generic conventions.
As remarked in the body of the paper such an analysis has been necessary in our case in order to obtain 
a 4d Lagrangian  suitable for the computation of the AdS$_2$ attractor in the formalism of \cite{DallAgata:2010ejj}.

We start from a 5d theory with $n_V$ vector multiplets and $n_H$ hypermultiplets and we indicate with an hat the fields that live in five dimensions. We use indices
\begin{align}
  \label{indices}
   I,J & = 1, \dots, n_V+1, & \Lambda,\Sigma & = 0,\dots, n_V+1, & u,v & = 1, \dots, 4n_H,\\
  \mu,\nu & = 1,\dots,4, & M,N & = 0,\dots, 4.
\end{align}
and we put hats on the five-dimensional quantities. We indicate the 5d vector fields as $\hat{A}^I_M$ and we use the same notations as in Appendix \ref{5dt}.\\
We reduce the five dimensional theory on a circle, using the following KK ansatz \cite{Duff:1986hr}:
\begin{equation}
  \label{KKansatz:metric}  
  \begin{split}
    d\hat{s}^2 & = e^{2\alpha\pt} ds^2 + e ^{2\beta\pt} (dy- A^0)^2,\\
    h^I & = -k \, e^{2\alpha\pt} z_2^I,\\
    \hat{A}^I_M & = \bigl( A^I_\mu -c \, z_1^I A_\mu^0, \, c \, z_1^I + \xi^I \bigr),
  \end{split}
\end{equation}
where $\alpha$, $\beta$, $c$ and $k$ are constants, $y$ is the direction of the circular fiber with range $4 \pi / g$, $A^0$ is the KK vector and $\xi^I$ are flat gauge connections. All the fields are independent of $y$. We are using the notation
\begin{equation}
  \label{reimz}
  z_1^I \equiv \Re \, z^I, \qquad z_2^I \equiv \Im \, z^I.
\end{equation}
Notice that in the last line of \eqref{KKansatz:metric} we are also performing a Scherk-Schwarz twist \cite{Benini:2020gjh, Hosseini:2017mds, Hristov:2014eba, Looyestijn:2010pb}. This is necessary to satify the 4d BPS equations and it will bring the extra Killing vector in the 4d theory.\\

Because of the constraint $\mathcal{V}(h)=1$ in \eqref{SM}, the field $\pt$ is redundant:
\begin{equation}
  \label{pt}
  e^{-6\alpha\pt} = -k^3 C_{IJK} z_2^I z_2^J z_2^K.
\end{equation}
We will use this relation during the computation.\\

We can fix the first constant by requiring that the dimensionally-reduced Lagrangian is of the Einstein-Hilbert form $\mathscr{L} = -\frac{1}{2} e  \, R +\dots$ In fact, the five-dimensional Ricci scalar reduces to
\begin{equation}
  \label{Ricci:5d}
  \hat{R} = e^{-2\alpha\pt} \bigl( R - 6\alpha^2 \partial_\mu \pt \partial^\mu \pt -\frac{1}{4} e^{-6\alpha \pt} F^0_{\mu\nu} F^{0 \, \mu\nu} \bigr) + \text{ total derivatives}
\end{equation}
and the determinant of the metric gives
\begin{equation}
  \label{e:5d}
  \hat{e} = e^{(\beta + 4\alpha)\pt} \, e. 
\end{equation}
Therefore, we have to impose $\beta = -2\alpha$.\\
We can thus rewrite the 5d metric and its inverse as
\begin{align}
  \label{metric:5d}
  \hat{g}_{MN} & =
  \begin{pmatrix}
    e^{2\alpha \pt} g_{\mu\nu} + e^{-4\alpha\pt} A^0_\mu A^0_\nu  & \qquad -e^{-4\alpha \pt} A^0_\mu \\
    -e^{-4\alpha \pt} A^0_\nu & e^{-4\alpha \pt}
  \end{pmatrix},\notag \\
  \hat{g}^{MN} & =
  \begin{pmatrix}
    e^{-2\alpha \pt} g^{\mu\nu} &  e^{-2\alpha\pt}A^{0\mu} \\
    e^{-2\alpha \pt} A^{0\nu} & \qquad e^{4\alpha\pt} + e^{-2\alpha \pt} A^0_\rho A^{0\rho}
  \end{pmatrix}, \notag \\
  \hat{e} & = e^{2\alpha\pt} e.
\end{align}
The reduction of the Einstein term follows from \eqref{Ricci:5d} and \eqref{e:5d}:
\begin{equation}
  \label{L1:5d}
  \hat{\mathscr{L}}_1 = \frac{\hat{e}\hat{R}}{2} = e \, \biggl( \frac{R}{2} - 3 \alpha^2 \partial_\mu \pt \partial^\mu \pt - \frac{e^{-6\alpha\pt}}{8} F^0_{\mu\nu} F^{0 \, \mu\nu} \biggr) + \text{ total derivatives}.
\end{equation} 
The reduction of the kinetic term of the scalars in the vector multiplet gives
\begin{equation}
  \label{L2:5d}
  \begin{split}
    \hat{\mathscr{L}}_2 & = - \hat{e} \, \frac{1}{2} g_{xy} \, \partial_M \phi^x \, \partial^M \phi^y = - \frac{3}{4} e \, a_{IJ} \, \hat{g}^{MN} \partial_M h^I \partial_N h^J \\
    & = -\frac{3k^2}{4} e \, a_{IJ} \, g^{\mu\nu} \partial_\mu \bigl( e^{2\alpha\pt} z_2^I \bigr) \, \partial_\nu \bigl( e^{2 \alpha \pt} z_2^J \bigr) \\
    & = e \, \Bigl[ -\frac{3k^2}{4} e^{4\alpha\pt} a_{IJ} \, \partial_\mu z_2^I \, \partial^\mu z_2^J + 3 \alpha^2 \, \partial_\mu \pt \partial^\mu \pt \, \Bigr],
  \end{split}
\end{equation}
where in the first line we used the relations \eqref{gxy:aIJ} and \eqref{hIx}, and the simplifications in the last line occur due to \eqref{hdh}, which implies
\begin{equation}
  \label{dx:dmu}
  h_I \, \partial_\mu \bigl[ h^I(\phi) \bigr] = h_I \, \partial_x h^I \partial_\mu \phi^x = 0,
\end{equation}
giving the condition on the 4d scalars
\begin{equation}
  \label{cond:zpt}
  2\alpha \, \partial_\mu \pt \, z_2^I + \partial_\mu z_2^I = 0.
\end{equation}
Notice that the last term in $\hat{\mathscr{L}}_2$ exactly cancels the second term in $\hat{\mathscr{L}}_1$. For simplicity we fix $\alpha=1$.\\
The reduction of the kinetic term of the scalars in the hypermultiplet gives
\begin{equation}
  \label{L3:5d}
  \begin{split}
    \hat{\mathscr{L}}_3 & = - \hat{e} \, \frac{1}{2} h_{uv} \, \hat{g}^{MN} \hat{\mathcal{D}}_M q^u \, \hat{\mathcal{D}}_N q^v\\
    & = e \, \Bigl[ -\frac{1}{2} h_{uv} \mathcal{D}_\mu q^u \mathcal{D}^\mu q^v - \frac{1}{2} e^{6\pt} g^2 (k_0^u + c \, z_1^I k_i^u)h_{uv} (k_0^v + c \, z_1^J k_J^v) \Bigr],
  \end{split}
\end{equation}
where $\hat{\mathcal{D}}_Mq^u$ is the 4d covariant derivative defined in \eqref{cov:q}, while
\begin{equation}
  \label{cov:q:4d}
  \mathcal{D}_\mu q^u = \partial_\mu q^u + g \, A_\mu^I k_I^u + g \, A_\mu^0 \xi^I k_I^u = \partial_\mu q^u + g \, A_\mu^\Lambda k_\Lambda^u
\end{equation}
is the 4d covariant derivative, which contains the extra Killing vector defined as
\begin{equation}
  \label{k0:4d}
  k_0^u := \xi^I k_I^u.
\end{equation}
The reduction of the gauge kinetic term gives
\begin{align}
  \label{L4:5d}
    \hat{\mathscr{L}}_4 & = -e \, \frac{1}{4} a_{IJ} \hat{F}^I_{MN} \hat{F}^{J \, MN}\\
    & = e \, \Bigl[ -\frac{1}{4} e^{-2\pt} a_{IJ} (F^I_{\mu\nu} - c \, z_1^I F^0_{\mu\nu} ) ( F^{J \mu\nu} - c \, z_1^J F^{0 \, \mu\nu}) - \frac{c^2}{2}e^{4\pt} a_{IJ} \, \partial_\mu z_1^I \, \partial^\mu z_1^J \, \Bigl] \notag,
\end{align}
where $\hat{F}_{MN}$ and $F_{\mu\nu}$ are the 5d and 4d field strengths. The first one can be obtained from the 5d vector fields in \eqref{KKansatz:metric}, i.e.
\begin{equation}
  \label{A:5d}
  \hat{A}^I = A^I - c \, z_1^I A^0 + (c \, z_1^I + \xi^I) dy.
\end{equation}
Thus
\begin{equation}
  \label{F:5d}
  \hat{F}^I = d\hat{A}^I = F^I - c \, dz_1^I \wedge A^0 - c \, z_1^I F^0 + c \, dz_1^I \wedge dy,
\end{equation}
from which we can read off the components
\begin{align}
  \label{F:5d:comp}
  \hat{F}^I_{\mu 4} = -\hat{F}^I_{4\mu} = c \, \partial_\mu z_1^I, \qquad \hat{F}^I_{\mu\nu} = F^I_{\mu\nu} -c \, z_1^I F^0_{\mu\nu} + c\, A_\mu^0 \partial_\nu z_1^I - c \, A_\nu^0 \partial_\mu z_1^I.
\end{align} 
To perform the reduction of the Chern-Simons term it is convenient to extend the geometry \eqref{KKansatz:metric} to a 6d bulk having the original 5d space as boundary. This can be obtained by extending the circle parametrized by the $y$ coordinate to a unit disk with radius $\rho \in [0,1]$ and the 5d connections $\hat{A}^I$ in \eqref{KKansatz:metric} to the following 6d ones:
\begin{equation}
  \label{A:6d}
  \tilde{A}^I = A^I + \xi^I A^0 + \rho^2 (z_1^I + \xi^I)(dy - A^0).
\end{equation}
We can thus rewrite the Chern-Simons action term as
\begin{equation}
  \label{CS:6d}
  \int_{5d} \hat{\mathscr{L}}_5 = \int_{5d} \frac{2}{3\sqrt{6}} \, C_{IJK} \hat{F}^I \wedge \hat{F}^J \wedge \hat{A}^K = \int_{6d} \frac{2}{3\sqrt{6}} \, C_{IJK} \tilde{F}^I \wedge \tilde{F}^J \wedge \tilde{F}^K,
\end{equation}
with $\tilde{F}^I = d\tilde{A}^I$. Integrating over $d\rho^2 \wedge (dy-A^0)$ we can extract the 4d Lagrangian:
\begin{equation}
  \label{L5:5d}
  \begin{split}
    \hat{\mathscr{L}}_5 & = \frac{1}{2\sqrt{6}} \, C_{IJK} \epsilon^{\mu\nu\rho\sigma} \Bigl[ ( c \, z_1^I + \xi^I ) F^J_{\mu\nu} F^K_{\rho\sigma} - ( c^2 \, z_1^I z_1^J - \xi^I \xi^J ) F^K_{\mu\nu} F^0_{\rho\sigma} \\
    & + \frac{ c^3 \, z_1^I z_1^J z_1^K + \xi^I \xi^J \xi^K}{3} F^0_{\mu\nu} F^0_{\rho\sigma} \Bigr].
  \end{split}  
\end{equation}
Finally, the reduction of the scalar potential gives
\begin{equation}
  \label{L6:5d}
  \begin{split}
    \hat{\mathscr{L}}_6 & = -\hat{e} \, g^2 \Bigl[ P_I^r P_J^r \bigl( 3 g^{xy} \partial_xh^I \partial_yh^J - 4 h^I h^J \bigr) + \frac{3}{4} h_{uv} k_I^u k_J^v h^I h^J \Bigr]\\
    & = -e^{2\pt} \, e \, g^2 \Bigl[ P_I^r P_J^r \bigl(3 g^{xy} \partial_xh^I \partial_yh^J - 4k^2 \, e^{4\pt} z_2^I z_2^J \bigr) + \frac{3k^2}{4} e^{4\pt} h_{uv}k_I^u k_J^v z_2^I z_2^J \Bigr]. 
  \end{split}
\end{equation}

We now rearrange the various pieces of the reduced Lagrangian to reproduce the general form of 4d $\mathcal{N}=2$ gauged supergravity with $n_V+1$ vector multiplets and $n_H$ hypermultiplets in \eqref{L:4d}.\\
The Einstein term descends from $\hat{\mathscr{L}}_1$:
\begin{equation}
  \label{L1:4d}
  \mathscr{L}_1 = \frac{e \, R}{2}.
\end{equation}
The kinetic term of the scalars in the vector multiplet receives contributions from $\hat{\mathscr{L}}_2$ and $\hat{\mathscr{L}}_4$:
\begin{equation}
  \label{L2:4d}
  \mathscr{L}_2 = -e \, e^{4\pt} a_{IJ} \Bigl( -\frac{3k^2}{4} \partial_\mu z_2^I \, \partial^\mu z_2^J + \frac{c^2}{2} \partial_\mu z_1^I \, \partial^\mu z_1^J \Bigr) = -e \, G_{I\bar{J}} \, \partial_\mu z^I \, \partial^\mu \bar{z}^{\bar{J}},
\end{equation}
where we defined the Hermitian metric
\begin{equation}
  \label{GIJ:4d}
  G_{I\bar{J}} := \frac{3k^2}{4} e^{4\pt} a_{I\bar{J}}.
\end{equation}
Notice that this recasting imposes a constraint between the parameters of the KK ansatz in \eqref{KKansatz:metric}, which we can fix as
\begin{equation}
  \label{c:k}
  c = \sqrt{\frac{3}{2}} k.
\end{equation}
The kinetic term of the scalars in the hypermultiplet is contained in $\hat{\mathscr{L}}_3$,
\begin{equation}
  \label{L3:4d}
  \mathscr{L}_3 = - \frac{e}{2} h_{uv} \mathcal{D}_\mu q^u \, \mathcal{D}^\mu q^v,
\end{equation}
where the covariant derivative $\mathcal{D}_\mu q^u$ is defined in \eqref{cov:q:4d} and \eqref{k0:4d}.\\
The scalar potential receives contributions from $\hat{\mathscr{L}}_3$ and $\hat{\mathscr{L}}_6$:
\begin{equation}
  \label{L6:4d:temp}
  \begin{split}
    \mathscr{L}_6 & = e^{-2\pt} \, g^2 \Bigl[ P_I^r P_J^r \bigl( 3 e^{2\pt} g^{xy} \partial_x h^I \partial_y h^J - 4k^2 e^{6\pt} z_2^I z_2^J \bigr) \\
    & + e^{6\pt} h_{uv} \bigl( \frac{3k^2}{4} k_I^u k_J^v z_2^I z_2^J + \frac{1}{2}( k_0^u + c \, z_1^I k_I^u )(k_0^v + c \, z_1^J k_J^v ) \bigr) \Bigr].
  \end{split}
\end{equation}
If we impose the constraint \eqref{c:k} and we fix
\begin{equation}
  \label{k}
  k = \sqrt{\frac{2}{3}},
\end{equation}
the second line of \eqref{L6:4d:temp} can be rewritten as
\begin{equation}
  \label{V:2}
  \frac{1}{2} e^{6\pt} h_{uv} k_\Lambda^u k_\Sigma^v X^\Lambda \bar{X}^\Sigma,
\end{equation}
in which we are using special coordinates $z^I = X^I/X^0$ in the K\"ahler frame $|X^0|^2=1$. (Recall that the indices $\Lambda, \Sigma$ run over 0 and then the values of $I,J$.) We will rearrange the first part of the scalar potential in a few lines.\\
Instead, we now focus on the gauge kinetic term, that gets contributions from $\hat{\mathscr{L}}_1$ and $\hat{\mathscr{L}}_4$:
\begin{equation}
  \label{L4:4d}
  \begin{split}
    \mathscr{L}_4 & = -e \, \frac{e^{-6\pt}}{8} \Bigl[ F^o_{\mu\nu} F^{0 \, \mu\nu} + 4 G_{IJ} \bigl( F^I_{\mu\nu} -  z_1^I F^0_{\mu\nu} \bigr) \bigl( F^{J \, \mu\nu} -  z_1^J F^{0 \, \mu\nu} \bigr) \Bigr] \\
    & = \frac{e}{8} \, \Im \, \mathcal{N}_{\Lambda\Sigma} \, F^\Lambda_{\mu\nu} F^{\Sigma \, \mu\nu}.
  \end{split}
\end{equation}
In the last line we recasted the field-dependent pieces in the matrix
\begin{equation}
  \label{ImN}
  \Im \, \mathcal{N}_{\Lambda\Sigma} = -e^{-6\pt}
  \begin{pmatrix}
    1+4 \, G_{MN} z_1^M z_1^N &\quad -4 \, G_{MJ} z_1^M \\
    -4 \, G_{IM}z_1^M & 4 \, G_{IJ}
  \end{pmatrix},
\end{equation}
which we will show later to be actually the imaginary part of the period matrix that descends from a proper prepotential.\\
Finally, we can rewrite $\hat{\mathscr{L}}_5$ as
\begin{equation}
  \label{L5:4d}
  \mathscr{L}_5 = \hat{\mathscr{L}}_5 = \frac{1}{12 \sqrt{6}} \Re \, \mathcal{N}_{\Lambda\Sigma} \epsilon^{\mu\nu\rho\sigma} F^\Lambda_{\mu\nu}F^\Sigma_{\rho\sigma},
\end{equation}
where
\begin{equation}
  \label{ReN}
  \Re \, \mathcal{N}_{\Lambda\Sigma} = \frac{4}{3\sqrt{6}}
  \begin{pmatrix}
    2\, C_{KLM} \bigl( z_1^K z_1^L z_1^M + \xi^K \xi^L \xi^M \bigr) & \ \ - 3 \, C_{JKL} \bigl( z_1^K z_1^L - \xi^K \xi^L \bigr) \\
    - 3 \, C_{IKL} \bigl( z_1^K z_1^L - \xi^K \xi^L \bigr) & 6 \, C_{IJK} \bigl( z_1^K + \xi^K \bigr)
    \end{pmatrix}.
\end{equation}
We now show that $G_{I\bar{J}}$ and $\mathcal{N}_{\Lambda\Sigma}$ come from the following prepotential:
\begin{align}
  \label{FX}
  F(X) & = \frac{4}{3 \sqrt{6}} C_{IJK} \frac{ \check{X}^I \check{X}^J \check{X}^K}{X^0} \qquad\qquad \text{with } \check{X}^I \equiv X^I + \xi^I X^0 \\
  & = \frac{4}{3 \sqrt{6}} C_{IJK} \biggl( \frac{X^I X^J X^K}{X^0}  + 3 \, \xi^I X^J X^K + 3 \, \xi^I \xi^J X^K X^0 + 2 \, \xi^I \xi^J \xi^K (X^0)^2 \biggr). \notag
\end{align}
Using special coordinates $z^I = X^I/X^0$, in the K\"ahler frame $|X^0|^2=1$, the K\"ahler potential \eqref{Kahler:pot} becomes
\begin{equation}
  \label{K:4d}
  \mathcal{K} = -\log \biggl( - \frac{32}{3\sqrt{6}} C_{IJK} z_2^I z_2^J z_2^K \biggr) = -\log \bigl( 8 e^{-6\pt} \bigr) .
\end{equation}
from which one can derive the K\"ahler metric \eqref{Kahler:metric}
\begin{equation}
  \label{G:4d}
  \begin{split}
    G_{i\bar{\jmath}} & = \sqrt{\frac{2}{3}} e^{6\pt} \Bigl( C_{IJK} z_2^K + \sqrt{\frac{2}{3}} e^{6\pt} C_{IKL}C_{JMN} z_2^K z_2^L z_2^M z_2^N \Bigr) \\
    & = 8 \sqrt{\frac{2}{3}} e^{\mathcal{K}} \Bigl( C_{IJK} z_2^K + 8 \sqrt{\frac{2}{3}} e^{\mathcal{K}} C_{IKL} C_{JMN} z_2^K z_2^L z_2^M z_2^N \Bigr),
  \end{split}
\end{equation}
that corresponds to \eqref{GIJ:4d} with \eqref{k}.\\
On the other hand, from the prepotential \eqref{FX} we obtain
\begin{equation}
  \label{F:LS}
  F_{\Lambda\Sigma} = \frac{4}{3\sqrt{6}}
  \begin{pmatrix}
    2 \, C_{KLM} \bigl( z^K z^L z^M + \xi^K \xi^L \xi^M \bigr) & \quad -3 \, C_{JKL} \bigl( z^K z^L - \xi^K \xi^L \bigr) \\
    -3 \, C_{IKL} \bigl( z^K z^L - \xi^K \xi^L \bigr) & 6 \, C_{IJK} \bigl( z^K + \xi^K \bigr)
  \end{pmatrix},
\end{equation}
from which we can derive the period matrix \eqref{periodmatrix}, whose real and imaginary parts are in agreement with \eqref{ReN} and \eqref{ImN}. In the computation of the period matrix we used the following relations:\\
\begin{align}
  \label{rel:1}
    (X^0)^{-2} X^\Lambda (\Im F_{\Lambda\Sigma}) X^\Sigma & = \frac{8}{3\sqrt{6}} C_{IJK} \Bigl( \Im ( z^I z^J z^K ) - 3 z^I \Im ( z^J z^K ) + 3 z^I z^J \Im(z^K) \Bigr) \notag \\
    & = - \frac{32}{3\sqrt{6}} C_{IJK} z_2^I z_2^J z_2^K = 8 e^{-6\pt} = e^{-\mathcal{K}},
\end{align}

\begin{equation}
  \label{rel:2}
  (X^0)^{-1} ( \Im F_{I\Sigma}) X^\Sigma = \frac{8i}{\sqrt{6}} C_{IKL} z_2^K z_2^L,
\end{equation}

\begin{equation}
  \label{rel:3}
   (X^0)^{-1} ( \Im F_{0\Sigma}) X^\Sigma = 2 e^{-6\pt} - \frac{8i}{\sqrt{6}} C_{IJK} z_1^I z_2^J z_2^K,
\end{equation}

\begin{equation}
  \label{rel:4}
  C_{IKL}C_{JMN} z_2^K z_2^L z_2^M z_2^N = \sqrt{\frac{3}{2}} e^{-6\pt} \Bigl(  \sqrt{\frac{3}{2}} e^{-6\pt}  G_{IJ} - C_{IJK} z_2^K \Bigr).
\end{equation}

Finally, we return to the 4d scalar potential in \eqref{L6:4d:temp}. To manipulate its first line, we notice that
\begin{equation}
  \label{inv:ImN}
  \bigl( \Im \, \mathcal{N} \bigr)^{-1 | \Lambda\Sigma} = -2 e^{\mathcal{K}}
  \begin{pmatrix}
    4 & 4 \, z_1^K \\
    4 \, z_1^J & \quad 4 \, z_1^J z_1^K + G^{JK}
  \end{pmatrix},
\end{equation}
and thus
\begin{equation}
  \label{inv+8}
  \bigl( \Im \, \mathcal{N} \bigr)^{-1 | \Lambda\Sigma} + 8 \, e^{\mathcal{K}} X^{(\Lambda}\bar{X}^{\Sigma)} = -2 \, e^{\mathcal{K}}
  \begin{pmatrix}
    0 & 0 \\
    0 & \quad G^{IJ}-4 \, z_2^I z_2^J
  \end{pmatrix}.
\end{equation}
Using this last relation, together with \eqref{aIJ}, the scalar potential becomes
\begin{equation}
  \label{L6:4d}
  \mathscr{L}_6 = -e \, g^2 \Bigl[ -\tilde{P}^r_\Lambda \tilde{P}^r_\Sigma \Bigl( \bigl(\Im \mathcal{N} \bigr)^{-1 | \Lambda \Sigma} + 8 \, e^{\mathcal{K}} X^\Lambda \bar{X}^\Sigma \Bigr) + 4 \, e^\mathcal{K} h_{uv} k_\Lambda^u k_\Sigma^v X^\Lambda \bar{X}^\Sigma \Bigr],
\end{equation}
where $\tilde{P}_\Lambda^r = 2 \, P_\Lambda^r$\footnote{The extra factor 2 is due to the difference of conventions between \cite{Benini:2020gjh} and \cite{Ceresole:2001wi}, see footnote \ref{P2P}.}\\
Notice that we cannot extract $\vec{P_0}$ directly from the potential, because it is multiplied by the null components of the matrix in \eqref{inv+8}. Nevertheless, being related to the Killing vector by \eqref{k:P}, it can be read from \eqref{k0:4d} and it is determined as $\vec{P_0} = \xi^I \vec{P_I}$.\\

The 4d Lagrangian that we have obtained performing this KK reduction exactly reproduces \eqref{L:4d}.\\

\bibliography{biblio}
\bibliographystyle{JHEP}

\end{document}